\documentclass[a4paper,11pt]{article}
\usepackage{jheppub}
\usepackage{overpic}

\pdfoutput=1
\usepackage{CJK}
\usepackage{xcolor}
\usepackage{amsfonts}
\usepackage{epstopdf}
\usepackage{wrapfig} 
\usepackage{subfigure}
\usepackage{graphicx}  
\usepackage{dcolumn}   
\usepackage{bm}
\usepackage{float}
\usepackage{lipsum}

\usepackage{epsfig}
\usepackage{psfrag}
\usepackage{amsmath}
\usepackage{amssymb}
\usepackage{color}
\usepackage{hyperref}

\newcommand{\be}{\begin{equation}}
\newcommand{\ee}{\end{equation}}
 \newcommand{\bea}{\begin{eqnarray}}
 \newcommand{\ena}{\end{eqnarray}}

\begin{document}

\newcommand\blfootnote[1]{%
\begingroup
\renewcommand\thefootnote{}\footnote{#1}%
\addtocounter{footnote}{-1}%
\endgroup
} 

\title{Formation and critical dynamics of topological defects in Lifshitz holography}

\author[\oplus1]{Zhi-Hong Li,}
\author[2]{Chuan-Yin Xia,$^\otimes$}
\author[3]{Hua-Bi Zeng,$^\dagger$}
\author[1,4]{Hai-Qing Zhang$^\ddagger$}
\blfootnote{$^{\oplus\otimes}$ZHL and CYX share the same contributions; $^{\dagger\ddagger}$Corresponding authors}



\affiliation[1]{Center for Gravitational Physics, Department of Space Science, Beihang University, Beijing 100191, China}
\affiliation[2]{Faculty of Science, Kunming University of Science and Technology, Kunming 650500, China}
\affiliation[3]{Center for Gravitation and Cosmology, College of Physics Science and Technology, Yangzhou University, Yangzhou 225009, China}
\affiliation[4]{International Research Institute for Multidisciplinary Science, Beihang University, Beijing 100191, China}

\emailAdd{lizhihong@buaa.edu.cn}
\emailAdd{chuanyinxia@foxmail.com}
\emailAdd{hbzeng@yzu.edu.cn}
\emailAdd{hqzhang@buaa.edu.cn}

\abstract{
We examine the formation and critical dynamics of topological defects via Kibble-Zurek mechanism in a (2+1)-dimensional quantum critical point, which is conjectured to dual to a Lifshitz geometry. Quantized magnetic fluxoids are spontaneously generated and trapped in the cores of order parameter vortices, which is a feature of type-II superconductors. Time evolution of the average condensate is found to lag behind the instantaneous equilibrium value of the order parameter, a typical phenomenon in non-equilibrium dynamics. Scalings of vortex number density and the ``freeze-out'' time match the predictions from Kibble-Zurek mechanism. From these scalings,  the dynamic and static critical exponents in the boundary field theory are found, at least at finite temperature, to be irrespective of the Lifshitz exponent in the bulk.}

\maketitle

\section{Introduction}
\label{sec:intro}
Critical phenomena are of great importance in modern physics and very few widely applicable principles are known for systems far-from equilibrium \cite{henkel,hohenberg}. In particular, understanding their critical dynamics in strongly coupled non-equilibrium phase transitions is extremely challenging \cite{Polkovnikov:2010yn}.  Among these, Kibble-Zurek mechanism (KZM) is a paradigmatic theory to describe the critical dynamics of the spontaneous generation of topological defects as the system undergoes a continuous phase transition \cite{Kibble:1976sj,Kibble:1980mv,Zurek:1985qw}. KZM has been tested and extended in various ways \cite{Chuang:1991zz,Bowick:1992rz,Baeuerle:1996zz,Ruutu:1995qz,Carmi:2000zz,Maniv:2003zz,Guo} (refer to \cite{Kibble:2007zz,delCampo:2013nla} for reviews). A continuous phase transition is characterized by the divergence of the coherence length $\xi$ and relaxation time $\tau$ near the critical point,
\begin{alignat}{1}\label{eq:9}
 \xi(\epsilon)=\xi_0|\epsilon|^{-\nu}, \qquad  \tau(\epsilon)=\tau_0|\epsilon|^{-\nu z_{\rm d}},
\end{alignat}
in which $\xi_0$ and $\tau_0$ are constant coefficients while $\nu$ and $z_{\rm d}$ are the static and dynamic critical exponents respectively.\footnote{We adopt $z_{\rm d}$ to denote the boundary dynamic critical exponent, in order to distinguish it from the Lifshitz exponent $z$ in the Lifshitz geometry.} $\epsilon$ is the dimensionless distance to the critical temperature: $\epsilon\equiv1-T/T_c$. KZM assumes a linear quench of the temperature from normal state to symmetry-breaking state with $\epsilon(t)=t/\tau_Q$, with $\tau_Q$ the quench time. At the ``freeze-out'' time $\hat t$, where the rate of change imposed from the quench is comparable to the relaxation time $\tau$, the system will adjust itself from nearly adiabatic to approximately impulse behavior, in which the order parameter effectively become ``frozen''. Thus, one can readily get
\begin{alignat}{1}\label{eq:tfreeze}
\hat t=(\tau_0\tau_Q^{\nu z_{\rm d}})^{\frac{1}{1+\nu z_{\rm d}}}.
\end{alignat}
Due to symmetry breaking, condensate of the order parameter will randomly distribute with each domain having the size $\hat \xi=\xi_0(\tau_Q/\tau_0)^{{\nu}/{(1+\nu z_{\rm d})}}$ and picks up their own constant phases. Topological defects thus will form at the vertex of some adjacent domains if their phases satisfy the so-called ``geodesic rule"\cite{Bowick:1992rz}.  Consequently, the resulting number density of defects in two-dimensional space can be estimated as
\begin{alignat}{1}\label{density}
 n\varpropto \hat \xi^{-2}=\xi_0^{-2}(\tau_0/\tau_Q)^{\frac{2\nu}{1+\nu z_{\rm d}}},
\end{alignat}
Eqs.\eqref{eq:tfreeze} and \eqref{density} are the central predictions of KZM.

 Gauge/gravity duality (AdS/CFT correspondence) is a ``first-principle'' means to study the strongly coupled field theories from weakly coupled gravitational theories in one higher dimensions \cite{Maldacena:1997re}. Applications of AdS/CFT in non-equilibrium dynamics can be found in, for instance \cite{Murata:2010dx,Garcia-Garcia:2013rha,MohammadiMozaffar:2019gpn}. For reviews, please refer to \cite{Liu:2018crr}. Previous holographic studies on KZM can be found in \cite{Chesler:2014gya,Sonner:2014tca,Zeng:2019yhi}. To find various scaling exponents in KZM has become a prime and important subject recently  \cite{delCampo:2013nla}. In this paper, we investigate the holographic KZM in the background of a Lifshitz geometry with various Lifshitz exponents $z$, which is conjectured to describe a quantum critical point on the boundary \cite{Kachru:2008yh,Hartnoll:2009ns,Taylor:2015glc}.
The quantized magnetic fluxes (fluxoids) are spontaneously generated and trapped in the cores of the order parameter vortices, a typical feature of type II superconductor. By investigating the scaling laws in Eqs.\eqref{eq:tfreeze} and \eqref{density}, we find that at least at finite temperature, the Lifshitz exponents $z$ in the bulk will not alter the dynamic critical exponents $z_{\rm d}$ in the boundary. In particular, it remains as $z_{\rm d}=2$.  This surprising conclusion is in line with previous studies in \cite{Natsuume:2018yrg,Evans:2010np} that at finite temperature, boundary field theory is like a mean field theory with $z_{\rm d}=2$, irrespective of the bulk Lifshitz exponent $z$.

\section{Holographic setup}
\label{sec:setup}
The gravity background we adopt is the $AdS_4$ black brane with Lifshitz exponent $z$ in Eddington-Finkelstein coordinates,
\begin{equation}\label{eq:1}
ds^{2}=-\frac{L^{2z}}{u^{2z}}f(u)dt^2-\frac{2L^{z+1}}{u^{z+1}}dtdu+\frac{L^2}{u^{2}}\left(dx^{2}+dy^{2}\right),
\end{equation}
where $f(u)=1-\left(u/u_{h}\right)^{2+z}$, $L$ is the AdS radius and $u$ is AdS radial coordinate. Location of the AdS boundary is $u=0$ while $u_h$ is the horizon. Hawking temperature of the black brane thus is $T=\frac{2+z}{4\pi L}\left({L}/{u_h}\right)^z$. Without loss of generality we rescale $L=u_h\equiv1$ in numerics.  The line element \eqref{eq:1} is invariant under the Lifshitz scaling
\begin{alignat}{1}\label{eq:2}
 t\rightarrow \lambda^z t;\qquad (u,x,y)\rightarrow \lambda (u,x,y);
\end{alignat}
The action we adopt is the commonly used Einstein-Maxwell-complex scalar action for holographic superconductors \cite{hartnoll}
\begin{eqnarray}\label{eq:3}
  S =\int d^4 x\sqrt{-g} \left[-\frac{1}{4}F_{\mu\nu}F^{\mu\nu}-|D\Psi|^2-m^2|\Psi|^2\right]
\end{eqnarray}
with $F_{\mu\nu}=\partial_{\mu}A_\nu-\partial_{\nu}A_\mu$, $D=\nabla-iA$.  We work in the probe limit by ignoring the backreaction of the matter fields to the gravitational fields.
The ansatz we take is $\Psi=\Psi(t,u,x,y),A_{t,x,y}=A_{t,x,y}(t,u,x,y)$ and $A_{u}=0$.
At the horizon, we demand the regularity of the fields.
Near the boundary $u\to0$, fields can be expanded as,
\begin{eqnarray}
  \Psi = \Psi_{0}u^{\Delta_{-}} + \Psi_{1}u^{\Delta_{+}},~~~
  A_{i}= a_i +b_iu^z, (i=x,y)
\end{eqnarray}
with $\Delta_{\pm}=\frac{z+2\pm\sqrt{(z+2)^2+4m^2}}{2}$ the conformal dimensions of dual scalar operators on boundary. The asymptotic behavior of $A_{t}$ is more sophisticated depending on $z$,
\begin{eqnarray}
A_t&=&a_t+b_t\log u,\qquad z=2, \label{z2expand}\\
A_t&=&a_t+b_tu^{2-z},\qquad z\neq2.
\end{eqnarray}
Following the AdS/CFT dictionary, $\Psi_{0}$, $a_t$ and $a_i$ are the source of the dual operator $O$, chemical potential and superfluid velocity in the boundary, respectively. Their corresponding conjugate variables can be achieved by varying the renormalized on-shell actions with respect to the source terms from holographic renormalization \cite{Skenderis:2002wp}.

\subsection{Holographic renormalization}
\label{sec:ren}
In order to get finite on-shell actions, counter terms should be added. For $z=1$, the counter term is $C_1=\int d^3x\sqrt{-\gamma}\left(n^\mu(D_\mu\Psi)^*\Psi+c.c.\right)$, where $\gamma$ is determinant of the reduced metric on the boundary while $n^\mu$ is the normal vector perpendicular to the boundary. In order to get the dynamical gauge fields in the boundary, we impose the Neumann boundary conditions for the gauge fields as $u\to0$ \cite{Witten:2003ya,Domenech:2010nf}. Thus, the surface term $C_{\rm surf.}=\int d^3x\sqrt{-\gamma}n^\mu F_{\mu\nu}A^\nu$  should also be added in order to have a well-defined variation. After doing these, one can get the finite renormalized on-shell action $S_{\rm ren.}$. Consequently, from the holographic renormalization we get the expectation value of the order parameter as $\langle O\rangle=\Psi_1$.
We also impose $\Psi_{0}=0$ in order to have the $U(1)$ symmetry spontaneously broken. Expanding the $u$-component of the Maxwell equations near boundary, we reach $\partial_tb_t+\partial_iJ^i=0$, which is exactly a conservation equation of the charge density $\rho$ and current $J^i$ on the boundary. Since from the variation of $S_{\rm ren.}$, one can get $b_t=-\rho$ and $J^i=-b_i-(\partial_ia_t-\partial_ta_i)$.
{{For $z=3/2$ the counter terms are similar to those of $z=1$. Consequently, near the boundary ($u\to0$) we get the conservation equation as $\partial_tb_t+\partial_iJ^i=0$ with $b_t=-\rho$ and $J^i=-3b_i/2-(\partial_ia_t-\partial_ta_i)$.}}
 For $z=2$ there are two counter terms $C_2=\int d^3x\sqrt{-\gamma}\Psi\Psi^*$ and $C_3=\int d^3x\sqrt{-\gamma}F_{ui}F^{ui}\log(\Lambda)$ with $\Lambda$ an ultraviolet cut-off. Besides, the above surface term $C_{\rm surf.}$ should also be added. Thus near $u\to0$ boundary we get the conservation equation as $\partial_tb_t+\partial_iJ^i=0$ with $b_t=-\rho$ and $J^i=-2b_i-(\partial_ia_t-\partial_ta_i)$.

\subsection{Numerical schemes}
\label{sec:num}

From the dimensional analysis in holographic superconductor \cite{hartnoll}, increasing the charge density equals decreasing the temperature. Therefore, we need to know the mass dimensions of the charge density for different Lifshitz exponents $z$. Following \cite{Hartnoll:2009ns}, this is usefully implemented by assigning time and space the following dimensions of mass $[t]=-z$, and  $[\vec{x}]=-1$. Thus, the temperature has dimensions $[T]=z$ and the charge density has mass dimension $[\rho]=2$. In order to linearize the temperature near the critical point according to KZM, we ought to quench the charge density $\rho$ as
\[\rho(t)=\begin{cases}\label{eq:density}
\rho_c(1-t/\tau_Q)^{-2}&\text{for $z=1$},\\
\rho_c(1-t/\tau_Q)^{-4/3}&\text{for $z=3/2$},\\
\rho_c(1-t/\tau_Q)^{-1}&\text{for $z=2$}.
\end{cases}\]
where $\rho_c$ is the critical charge density for the static and homogeneous holographic superconducting system.

In this paper, we choose the Lifshitz exponents $z$ in the bulk as {{$z=(1,3/2,2)$}}. Physically, this corresponds to relativistic and non-relativistic systems in the boundary, respectively. Besides, we would like to investigate the properties of dual scalar operators with the same conformal dimension, for convenience we set $\Delta_{+}=3$ as we vary $z$. {{Therefore, the mass squares are $m^2=(0,-3/2,-3)$ with respect to $z=(1,3/2,2)$. Correspondingly, the critical charge densities for the static homogeneous superconductors are $\rho_c\approx 7.5877$ for $z=1$, $\rho_c\approx 8.4278$ for $z=3/2$ and $\rho_c\approx 9.0445$ for $z=2$.}}

We take advantage of the Chebyshev pseudo-spectral method with 21 grids in the radial direction $u$ and use the Fourier decomposition in the $(x,y)$-directions since the periodic boundary condition along $(x,y)$ was imposed.  We thermalize the system by adding small random seeds in the normal state before quench. The reason is to make sure that the system before quench is in a symmetrical phase, which is the requirement of KZM.  Different from putting the seeds on the boundary in \cite{Chesler:2014gya,Sonner:2014tca}, we add the random seeds of the fields in the bulk by satisfying the distributions $\langle s(t,x_\mu)\rangle=0$ and $\langle s(t,x_\mu)s(t',x'_\mu)\rangle=\zeta\delta(t-t')\delta(x_\mu-x'_\mu)$ where $(\mu=u,x,y)$, with the amplitude  $\zeta\approx10^{-3}$. \footnote{Other relatively smaller magnitudes of $\zeta$ lead to similar results. In principle, $\zeta$ cannot be too large since the seeds play the role of perturbations to thermalize the system.}  The system evolves by using the fourth order Runge-Kutta method with time step $\Delta t=0.02$ for $z=1$, {{$\Delta t=0.01$ for $z=3/2$}} and $\Delta t=0.0046$ for $z=2$. Filtering of the high momentum modes are implemented following the ``$2/3$'s rule'' that the uppermost one third Fourier modes are removed \cite{Chesler:2013lia}.

\begin{figure}[h]
\centering
\includegraphics[trim=3.6cm 9.8cm 3.8cm 10.cm, clip=true, scale=0.5,  angle=0] {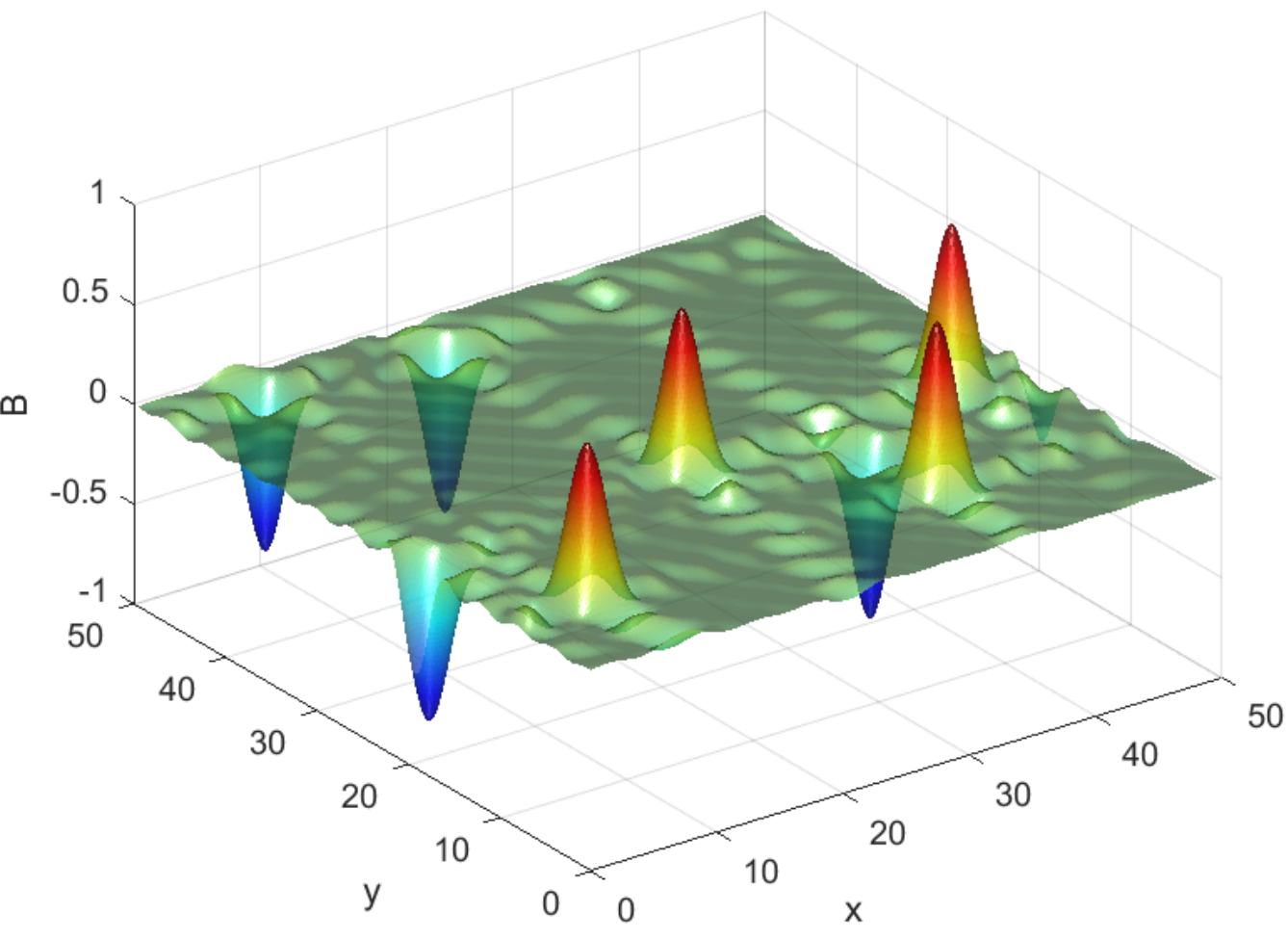}
~~~
\includegraphics[trim=3.8cm 9.8cm 3.8cm 10.cm, clip=true, scale=0.5,  angle=0] {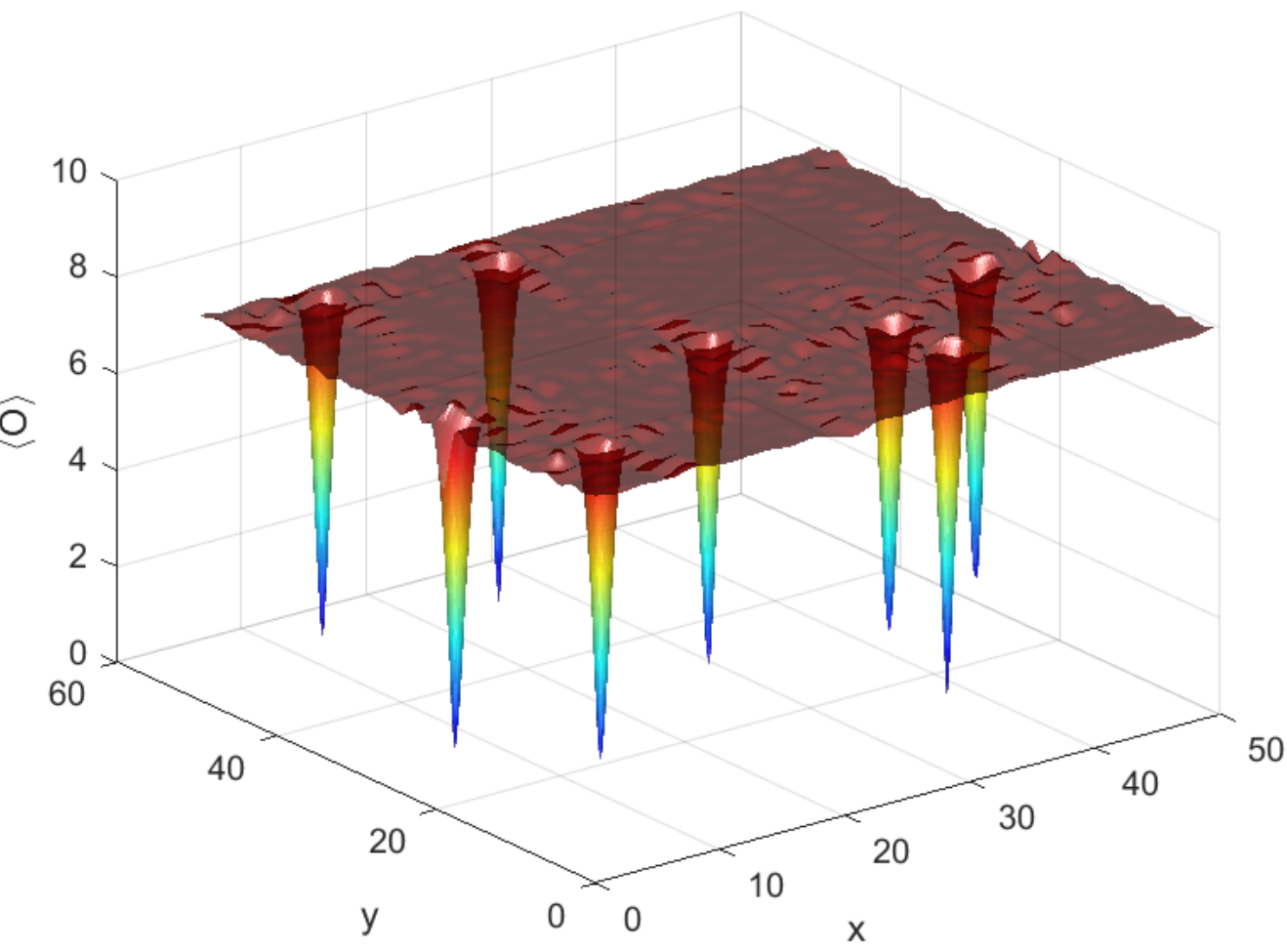}
\caption{\label{BO}Configurations of eight (four positive and four negative) magnetic fluxoids (left panel) and superconducting vortices (right panel) at temperature $T=0.8T_c$ with the quench time $\tau_Q=1800$ for Lifshitz exponent $z=2$. Locations of the order parameter vortices exactly correspond to the positions of the magnetic fluxoids.}
\end{figure}

\section{Numerical results}
\subsection{Magnetic fluxoids and order parameter vortices}
\label{sec:fluxon}

We quench the system by linearly decreasing the temperature through the critical point, then stop and keep the temperature at $T=0.8T_c$. $t=0$ is the instant to cross the critical temperature $T_c$.
In the left panel of Fig.\ref{BO}, we show the magnetic fluxes generated from KZM as the system enters the final equilibrium state with $\tau_Q=1800$ and Lifshitz exponent $z=2$. In the right panel of Fig.\ref{BO}, the corresponding order parameter vortices are exhibited. The locations of the cores of the vortices are exactly the positions of the magnetic fluxes, which is a feature of type II superconductor.

\begin{figure}[h]
\centering
\includegraphics[trim=1.3cm 7.9cm 1.5cm 6.5cm, clip=true, scale=0.25,  angle=0] {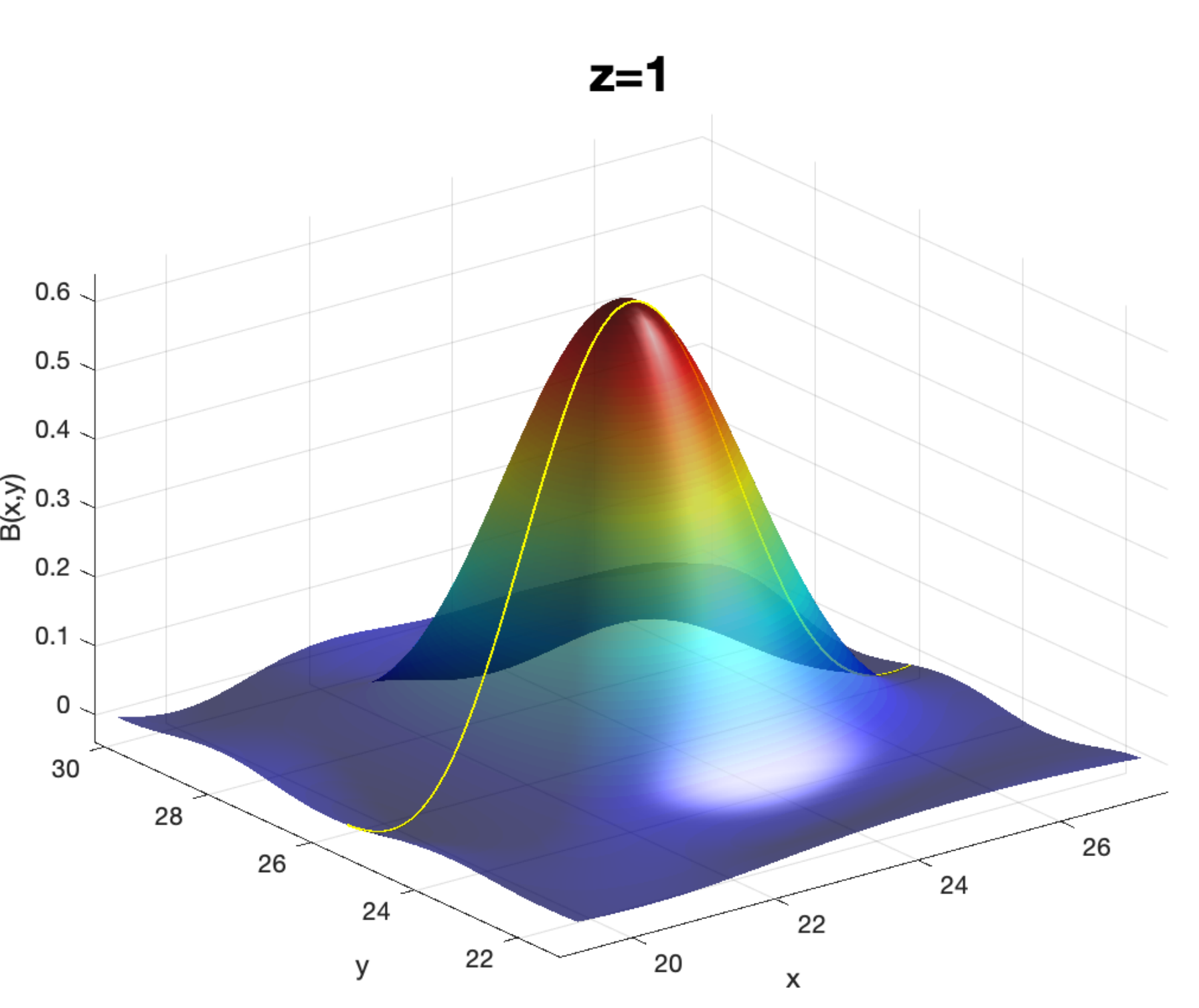}
~
\includegraphics[trim=1.3cm 7.9cm 1.5cm 6.5cm, clip=true, scale=0.25,  angle=0] {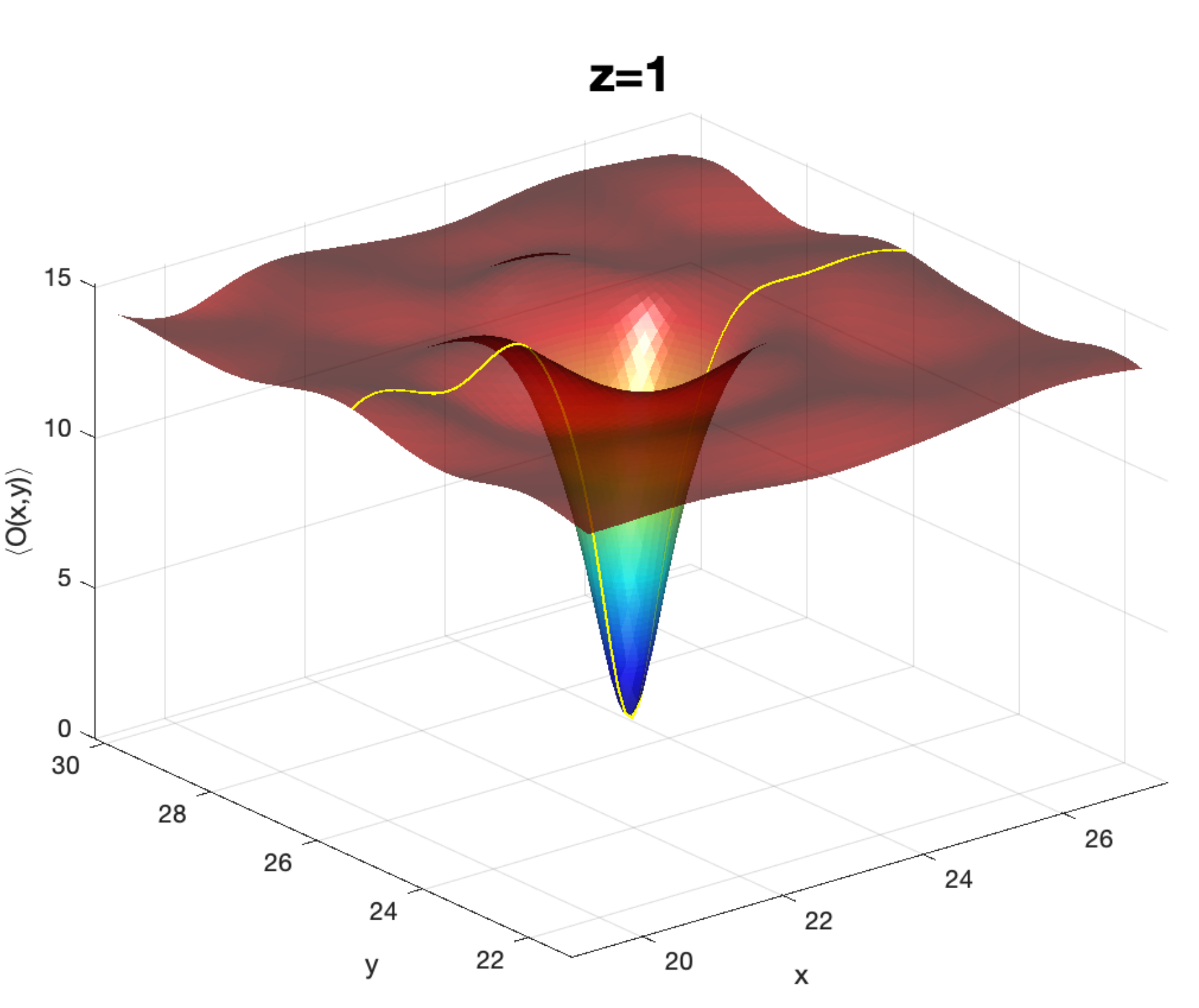}
~
\includegraphics[trim=2cm 6.5cm 1cm 6.5cm, clip=true, scale=0.26,  angle=0] {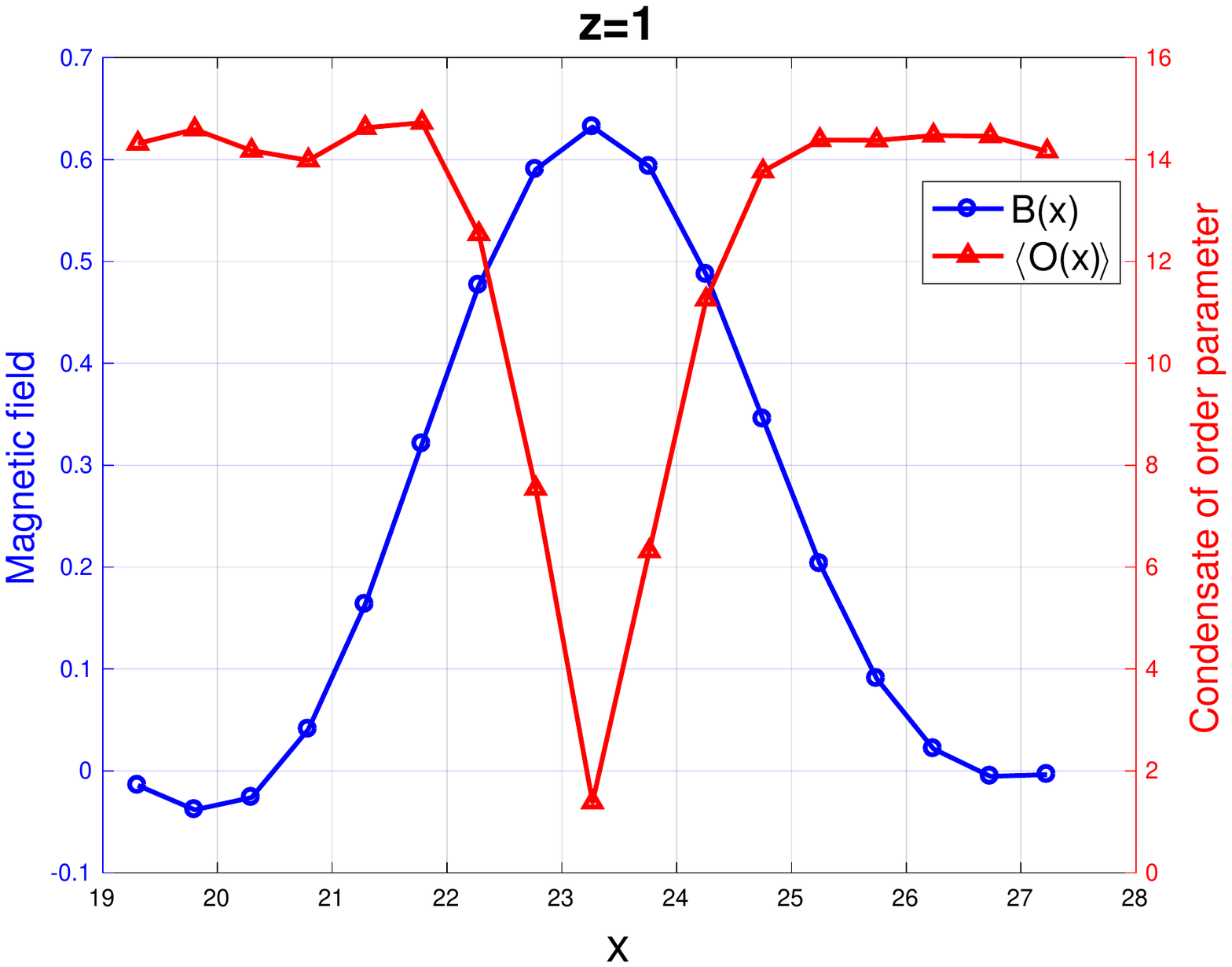}
\includegraphics[trim=3.5cm 6.9cm 3cm 9.5cm, clip=true, scale=0.30,  angle=0] {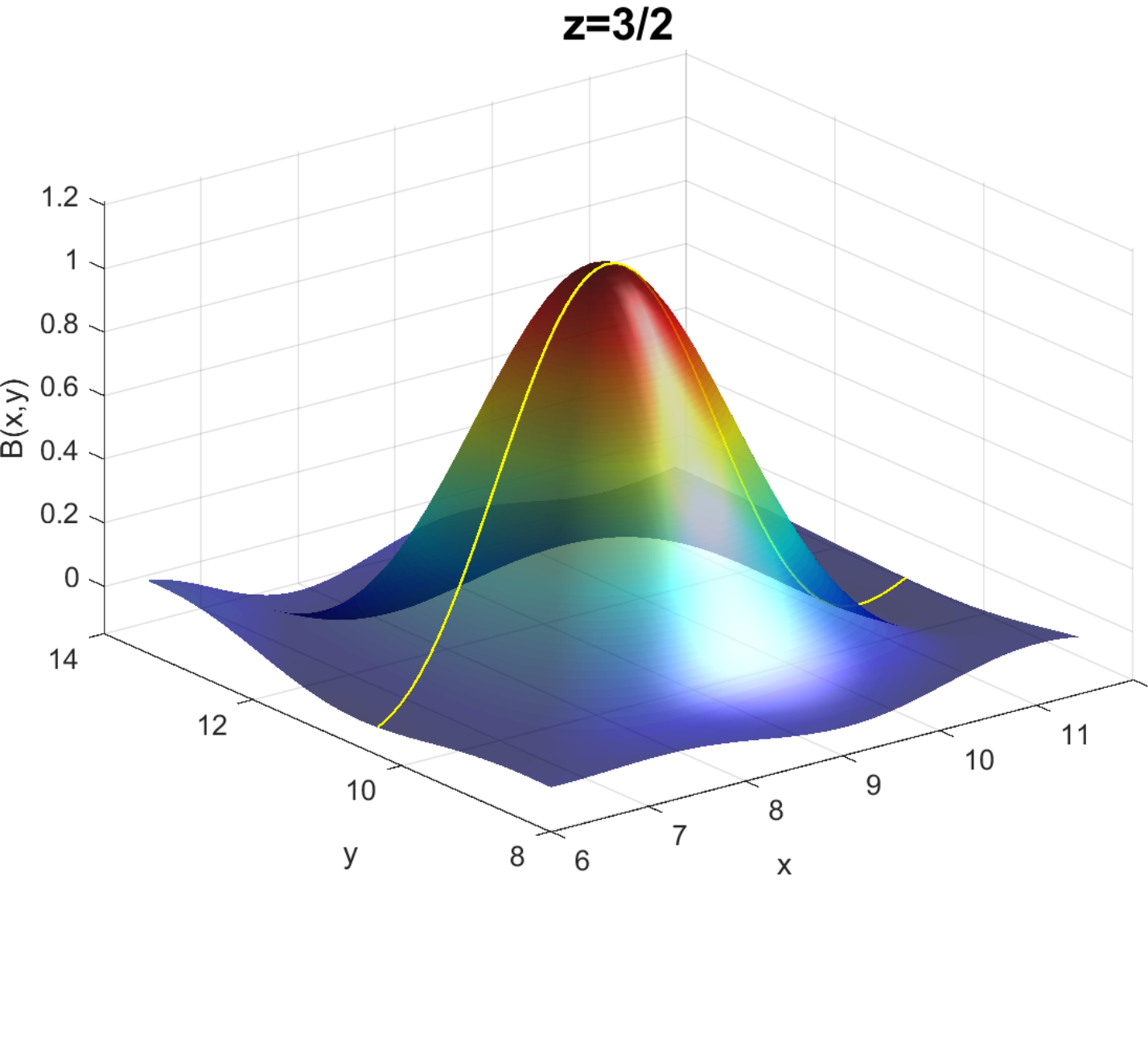}
\includegraphics[trim=3cm 6.6cm 2cm 10.6cm, clip=true, scale=0.315,  angle=0] {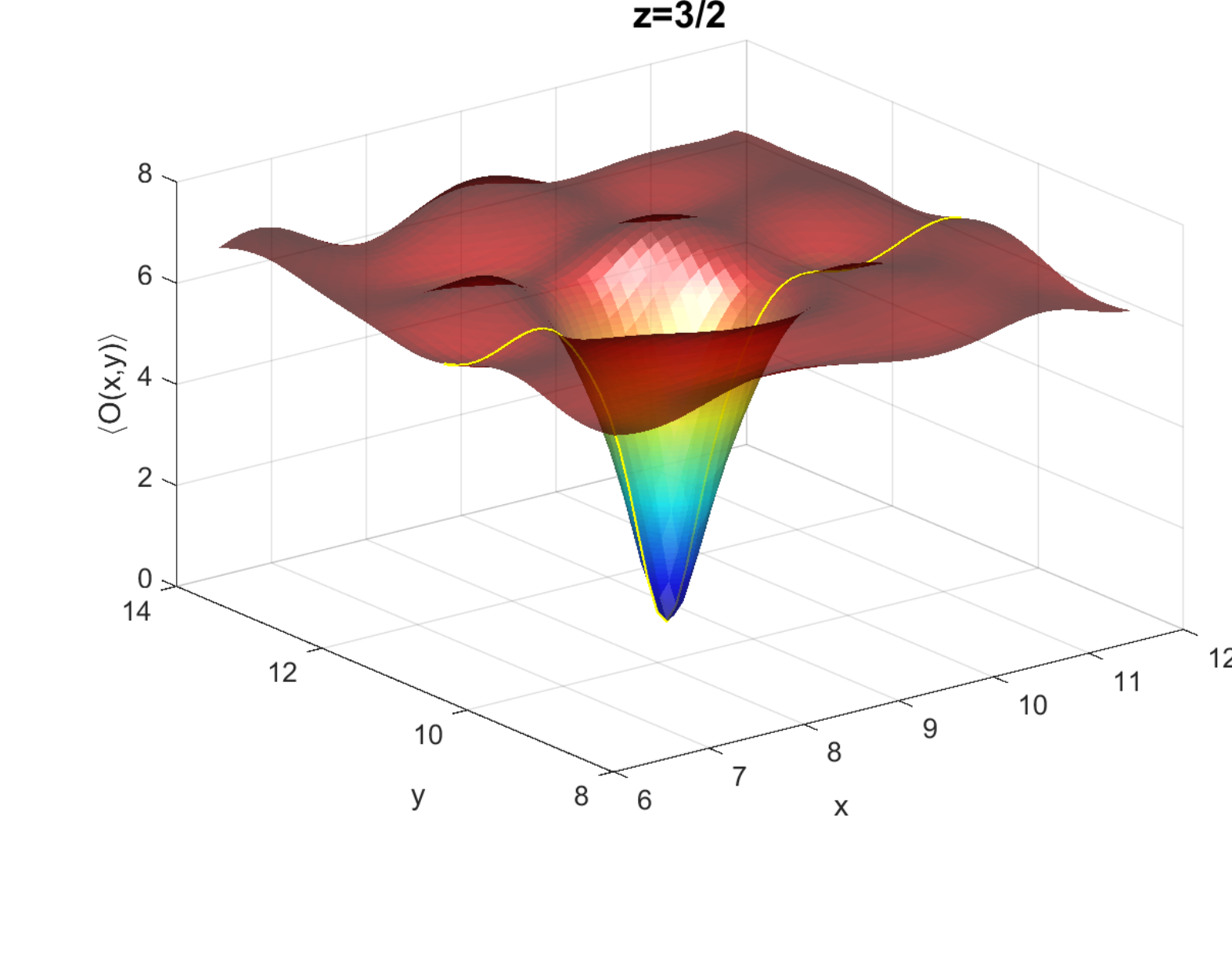}
\includegraphics[trim=3.3cm 6.5cm 1.8cm 7.5cm, clip=true, scale=0.33,  angle=0] {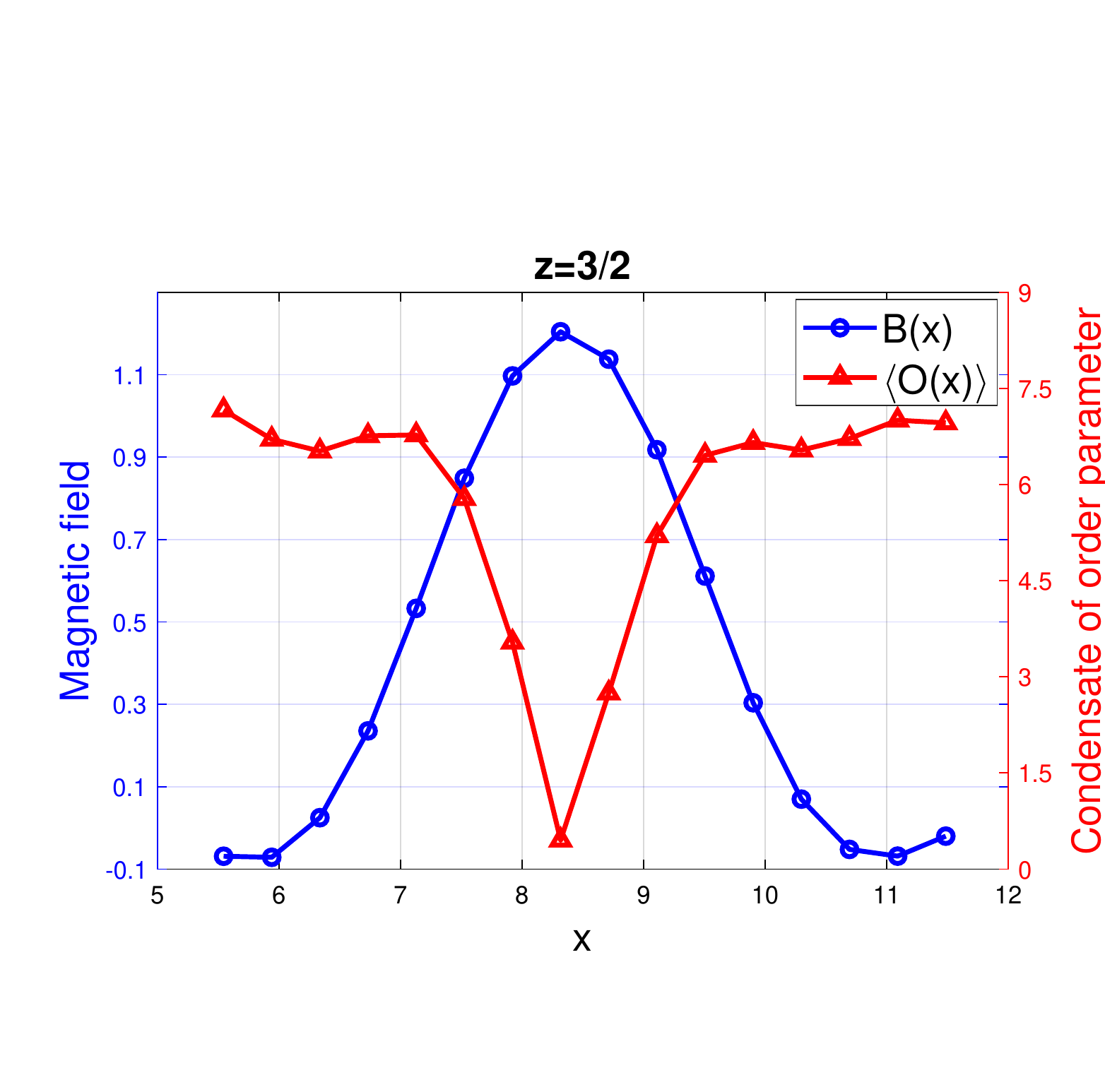}
\includegraphics[trim=1.3cm 7.9cm 1.5cm 6.5cm, clip=true, scale=0.25,  angle=0] {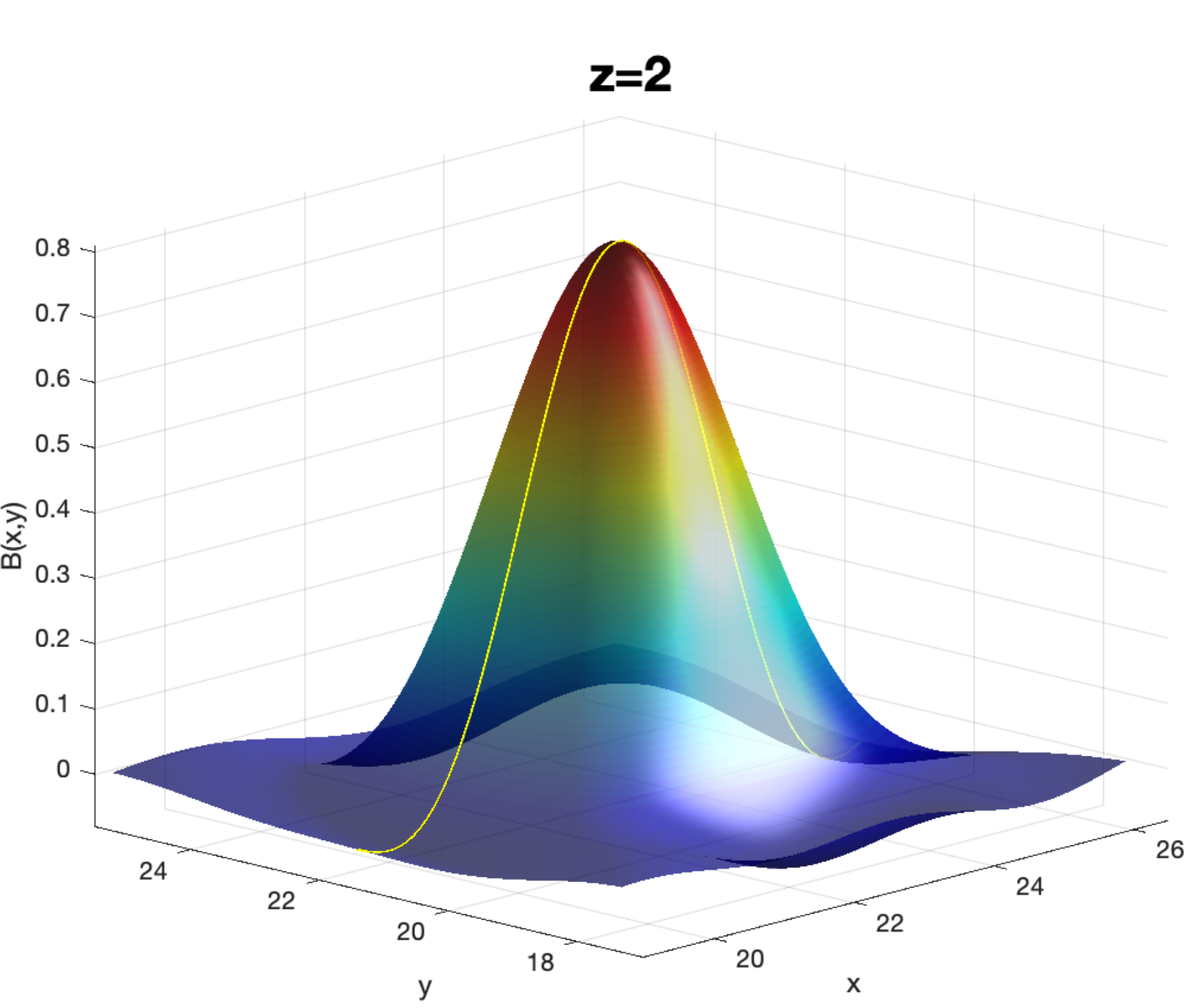}
~
\includegraphics[trim=1.3cm 7.9cm 1.5cm 6.5cm, clip=true, scale=0.25,  angle=0] {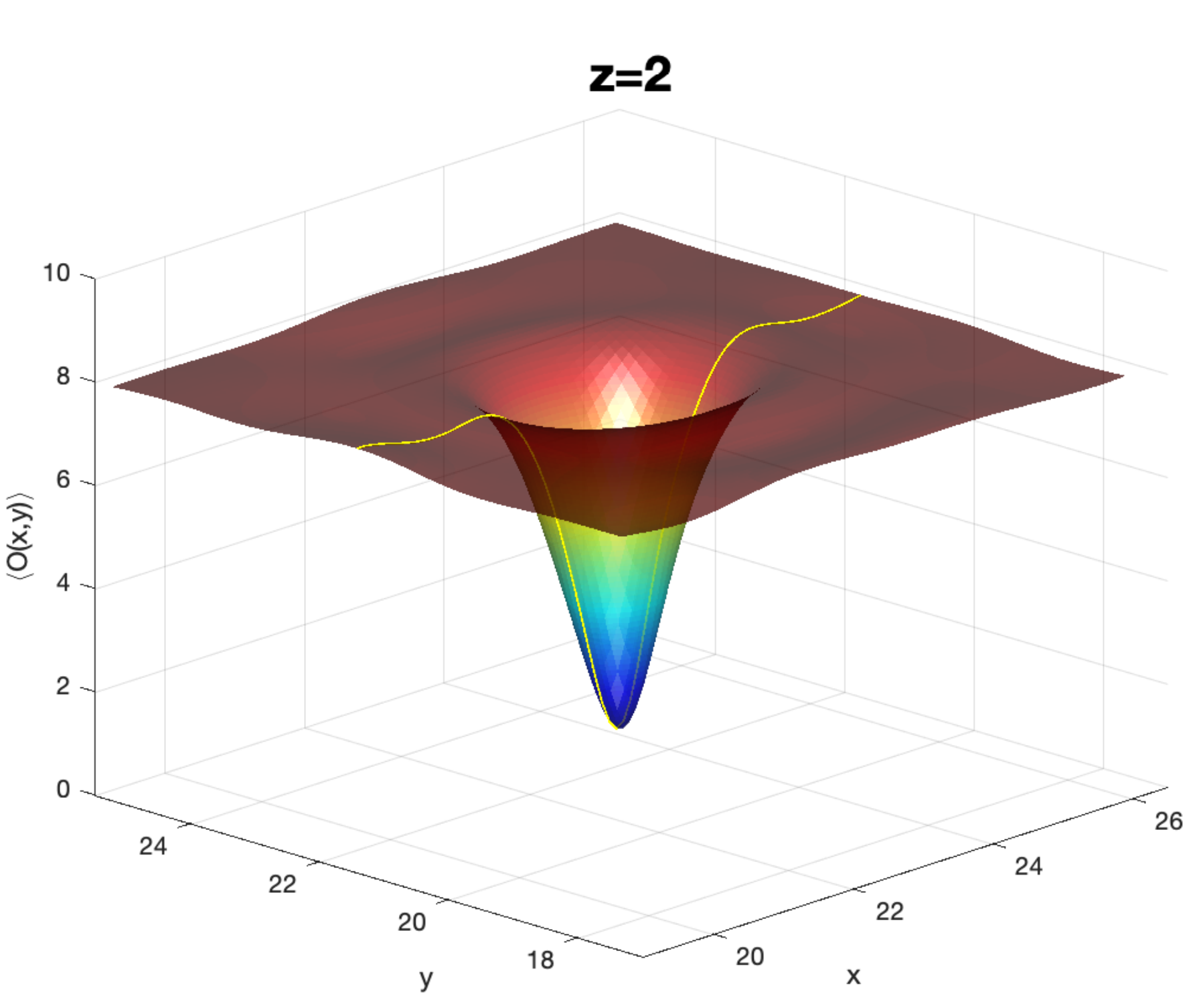}
~
\includegraphics[trim=2cm 6.5cm 1cm 6.5cm, clip=true, scale=0.25,  angle=0] {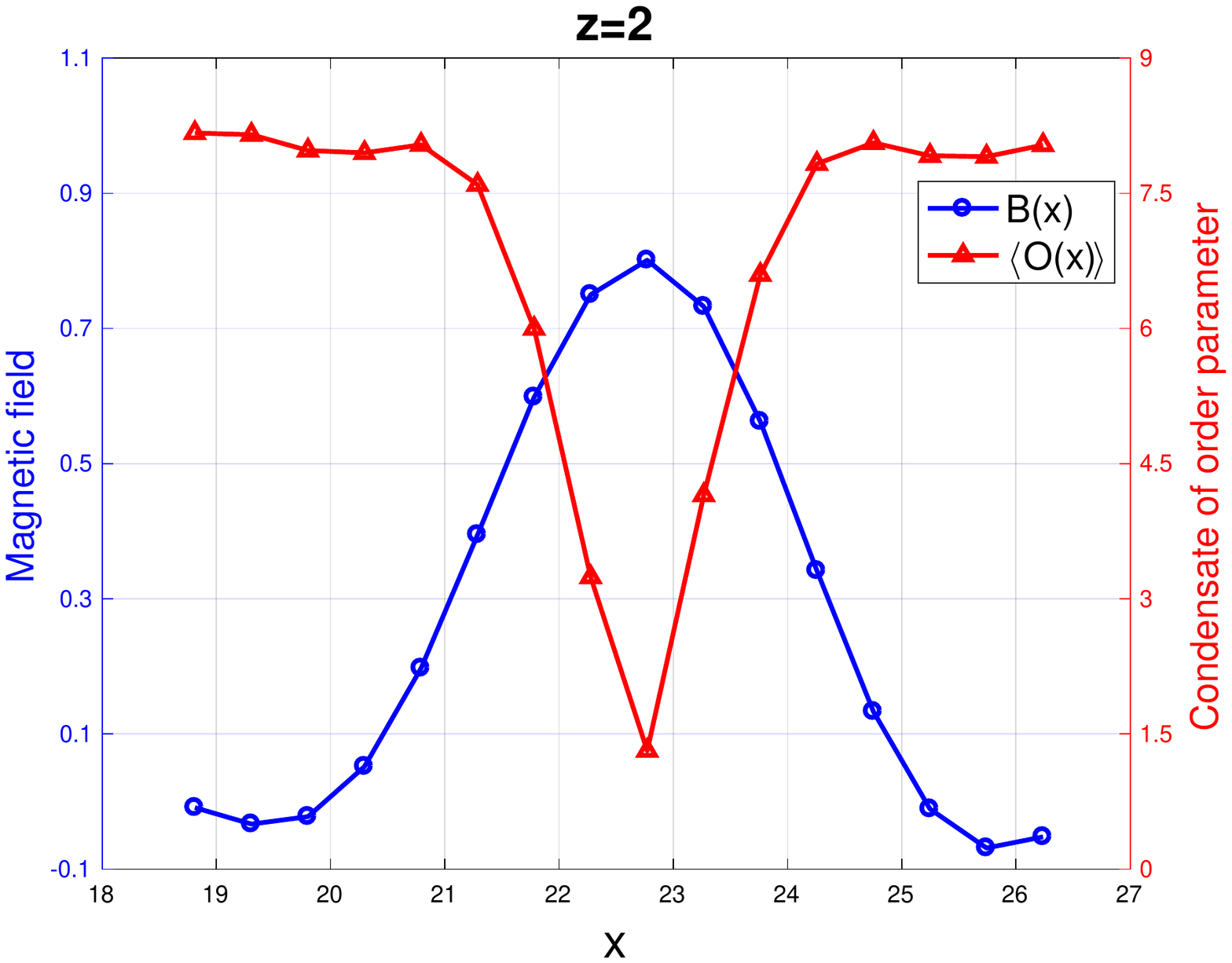}
\caption{\label{1dBO}
{{Three-dimensional configurations of the single magnetic fluxoid and the order parameter vortex for $z=1$ (top row), $z=3/2$ (middle row) and $z=2$ (bottom row).}} Yellow lines are the cross sections along $x$-direction. Their corresponding data are shown in the last column.  From the widths of the magnetic fluxoids (blue lines) and the order parameter vortices (red lines), it is found that they are all type II superconductors.  }
\end{figure}

The profiles of a single vortex can be seen in Fig.\ref{1dBO}, in which we select the vortex at the location $(x,y)\approx(23,26)$ for $z=1$ (top row), {{$(x,y)\approx(9,11)$ for $z=3/2$ (middle row) }} and $(x,y)\approx(23,22)$ for $z=2$ (bottom row). Requirements of minimal free energy and periodicity of the phase of order parameter imply the quantization of magnetic flux \cite{tinkham}, i.e., {\it magnetic fluxoid} with flux $\Phi_c=2\pi N$ where $N$ is an integer. By integrating the magnetic flux of the single vortex numerically for  {{$z=(1, 3/2, 2)$ from top to bottom of Fig.\ref{1dBO}, we find the magnetic flux for this vortex is $\Phi_c\approx(6.11,6.14, 6.18)$, }} which demonstrates the existence of quantized magnetic fluxoid with $\Phi_c=2\pi$ (winding number $N=1$). Other magnetic fluxes, for instance of vortices in Fig.\ref{BO}, are also checked to be quantized with $N=\pm1$ vorticity.

The width of the magnetic field $\lambda$ can be fitted by $B(r)\sim B_0 e^{-r/\lambda}$ with $B_0$ constant coefficient, while the width of the order parameter vortex $w$ is fitted by $O(r)\sim O(\infty)\tanh(r/(\sqrt{2}w))$ with $O(\infty)$ the condensate value far from the vortex core \cite{tinkham}. From the the last column in Fig.\ref{1dBO}, we find $\lambda\approx 1.69$ and $w\approx0.88$ for $z=1$. Thus, the Landau-Ginzburg parameter $\kappa_{z=1}=\lambda/w\approx1.92>1/\sqrt{2}$, which indicates a type II superconductor. {{For $z=3/2$ and $z=2$ we obtain the similar conclusion of the type II superconductor with $\kappa_{z=3/2}=\lambda/\xi=1.10/0.71=1.55>1/\sqrt{2}$ and $\kappa_{z=2}=\lambda/\xi=1.35/0.75=1.8>1/\sqrt{2}$. }} These results of type II superconductors are consistent with the appearance of the magnetic fluxoids in a holographic superconductor.

\begin{figure}
\centering
\includegraphics[trim=3.5cm 9.0cm 3.5cm 9.2cm, clip=true, scale=0.37, angle=0] {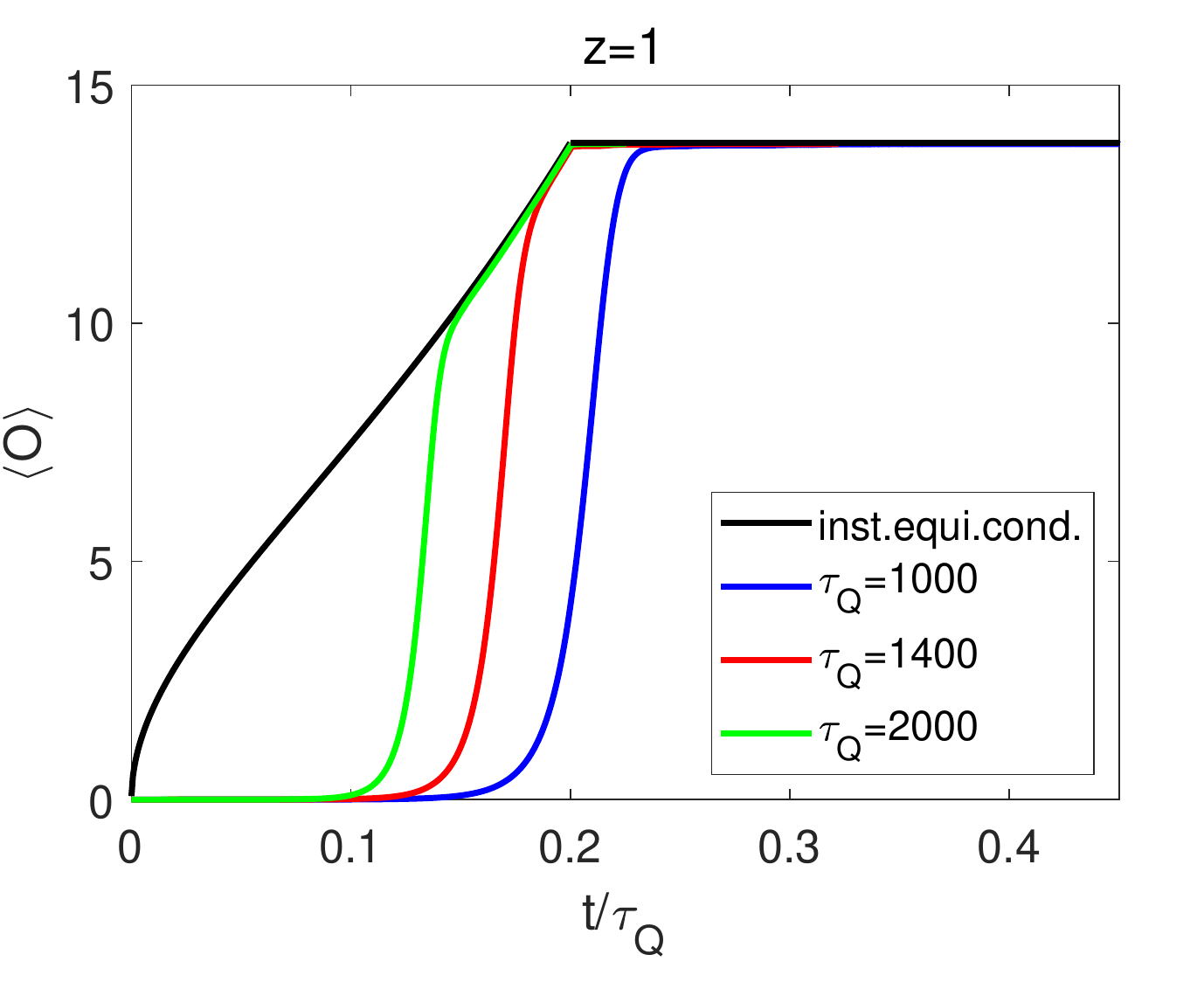}
\includegraphics[trim=2.4cm 7.8cm 2.6cm 8.7cm, clip=true, scale=0.315, angle=0] {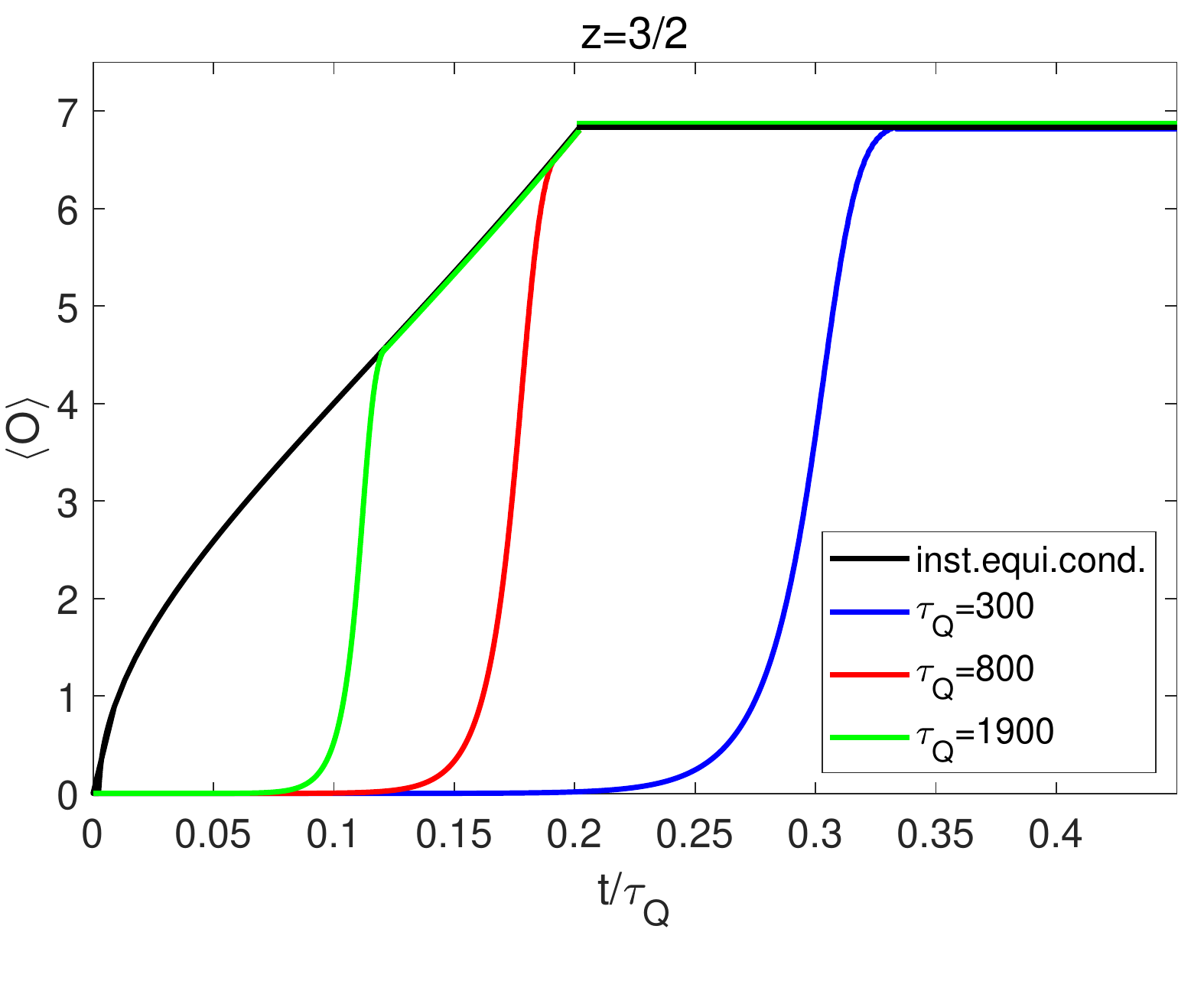}
\includegraphics[trim=3.5cm 9.0cm 3.5cm 9.2cm, clip=true, scale=0.37, angle=0] {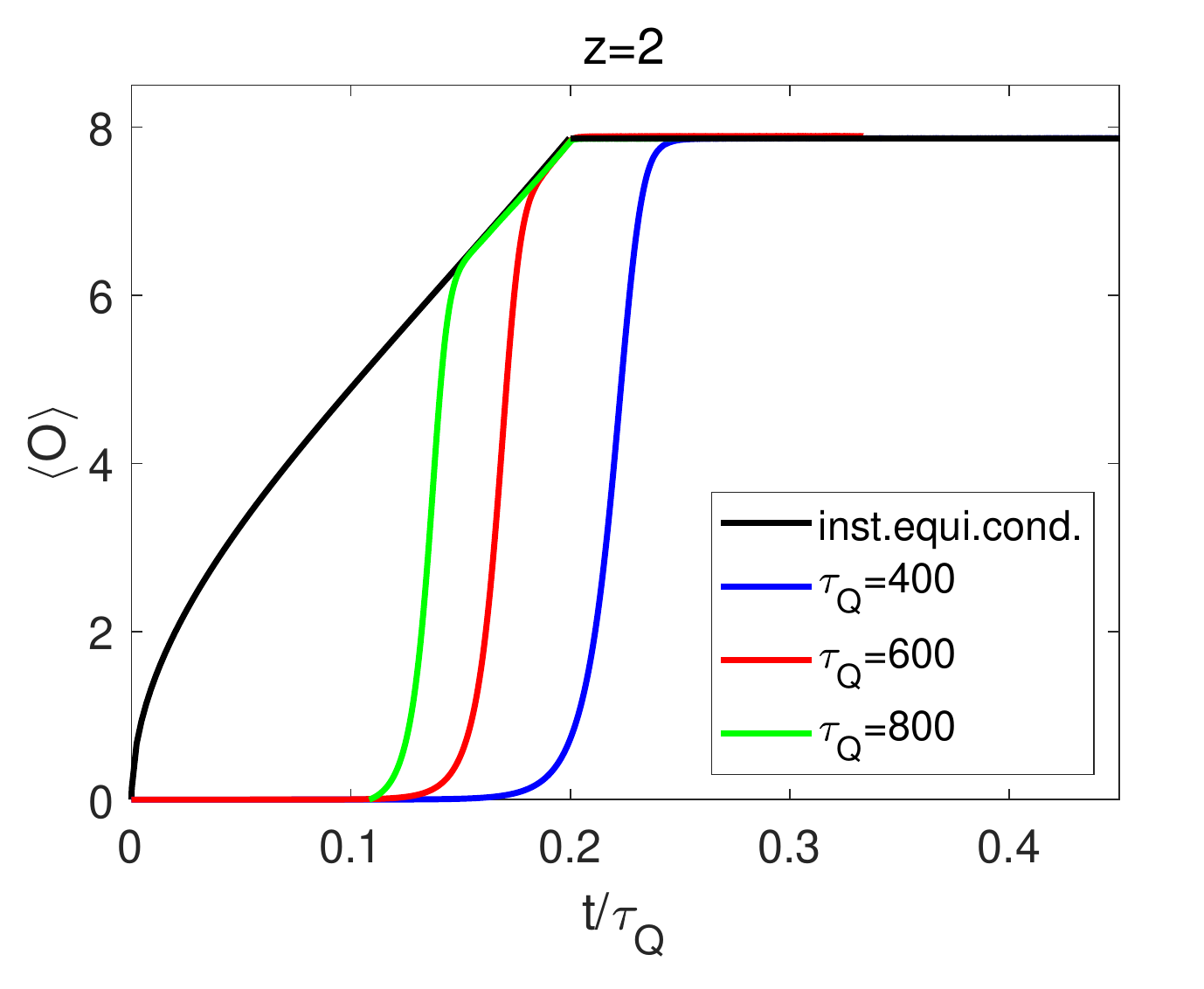}
\caption{\label{deltaT2} Time evolution of the average condensate with Lifshitz exponent $z=1$ (left panel),$z=3/2$ (middle panel) and $z=2$ (right panel) for various $\tau_Q$'s. Black lines stand for the instantaneous equilibrium values of the condensate. Colored lines are the dynamical values of the condensate under different quenches. For all panels, the final temperatures are $T=0.8T_c$ and $t=0$ is the instant to cross the critical point.}
\end{figure}

\subsection{Evolution of average condensate}
\label{sec:evolution}
Time evolution of the average value of order parameter $\langle O(t) \rangle$ from $t=0$ ($T=T_c$) to the final equilibrium state ($T=0.8T_c$) is exhibited in Fig.\ref{deltaT2}.  In the left panel with Lifshitz exponent $z=1$, the black line is the instantaneous equilibrium value of the average condensate, while the colored lines from left to right correspond to the dynamical values of the average condensate under different quenches $\tau_Q=2000 (\rm green)$, $1400 (\rm red)$ and $1000(\rm blue)$, respectively. {{Explanations for lines in the panels with $z=3/2$ and $z=2$ are direct from the figures}}.

From Fig.\ref{deltaT2}, we see that in the beginning the dynamical values of condensate remain negligible, and lags behind the instantaneous equilibrium values. For instance of quench $\tau_Q=1000$ in the left panel ($z=1$), its dynamical value remains negligible until the lag time $t_L/\tau_Q\sim0.15$, and then begins to scramble rapidly reaching the approximate equilibrium value at $t/\tau_Q\sim0.23$.  This behavior, with $t_L$ larger than but proportional to $\hat t$,  was reported as well in previous literatures \cite{Sonner:2014tca,Zeng:2019yhi,Das:2011cx}. For convenience, we list in Table.\ref{table} the approximate values of $t_L$ for various quenches presented in Fig.\ref{deltaT2}. From Table.\ref{table} we see that for the same Lifshitz exponent, slower quench (bigger $\tau_Q$) corresponds to longer lag time $t_L$. This is consistent with Eq.\eqref{eq:tfreeze} if the power $\nu z_d/(1+\nu z_d)$ is positive. We will further discuss the relation between $t_L$ and $\tau_Q$ in the next subsection, and indeed we will see there that Table.\ref{table} is consistent with Eq.\eqref{eq:tfreeze}.

\begin{center}
\begin{table}[h]
\centering
\begin{tabular}{|c|c|c|c||c|c|c||c|c|c|}
    \hline \multicolumn{4}{|c||}{$z=1$}&\multicolumn{3}{c||}{$z=3/2$}&\multicolumn{3}{c|}{$z=2$}\\
    \hline
    $\tau_Q$ & $1000$ & $1400$ & $2000$&300&800&1900 &400&600&800 \\
    \hline
    $t_L$ & 150 & 175 & 200 &71&114&173&70&80&90 \\
    \hline
    $t_L/\tau_Q$ & 0.15 & 0.125 & 0.1 &0.235&0.14&0.09&0.175&0.133&0.1125  \\
    \hline
  \end{tabular}
     \caption{Approximate values of $t_L$ under various quench times $\tau_Q$ for $z=(1, 3/2, 2)$.}
    \label{table}
    \end{table}
\end{center}

In addition, we see that for the same Lifshitz exponent, the final equilibrium condensates under different $\tau_Q$'s are almost identical since the final temperatures are the same ($T=0.8T_c$).  Another interesting phenomenon is that for slower quench, for instance of $\tau_Q=2000$ in the left panel of Fig.\ref{deltaT2} ($z=1$), after the lag time the dynamical condensate rapidly grows, then catches up and coincides with the instantaneous equilibrium value (black line). This behavior was also reported in the past \cite{Sonner:2014tca,Chesler:2014gya,Zeng:2019yhi}. The end of this coincident growth ($t/\tau_Q=0.2$) exactly corresponds to the end of quench, and the coincidence indicates an adiabatic region for the growth. However, for the fast quench (for example $\tau_Q=1000$ with $z=1$) there is no such kind of coincidence of the condensate before the end of quench.

\begin{figure}[h]
\centering
\includegraphics[trim=0cm 0cm 0cm 0cm, clip=true, scale=0.38,  angle=0] {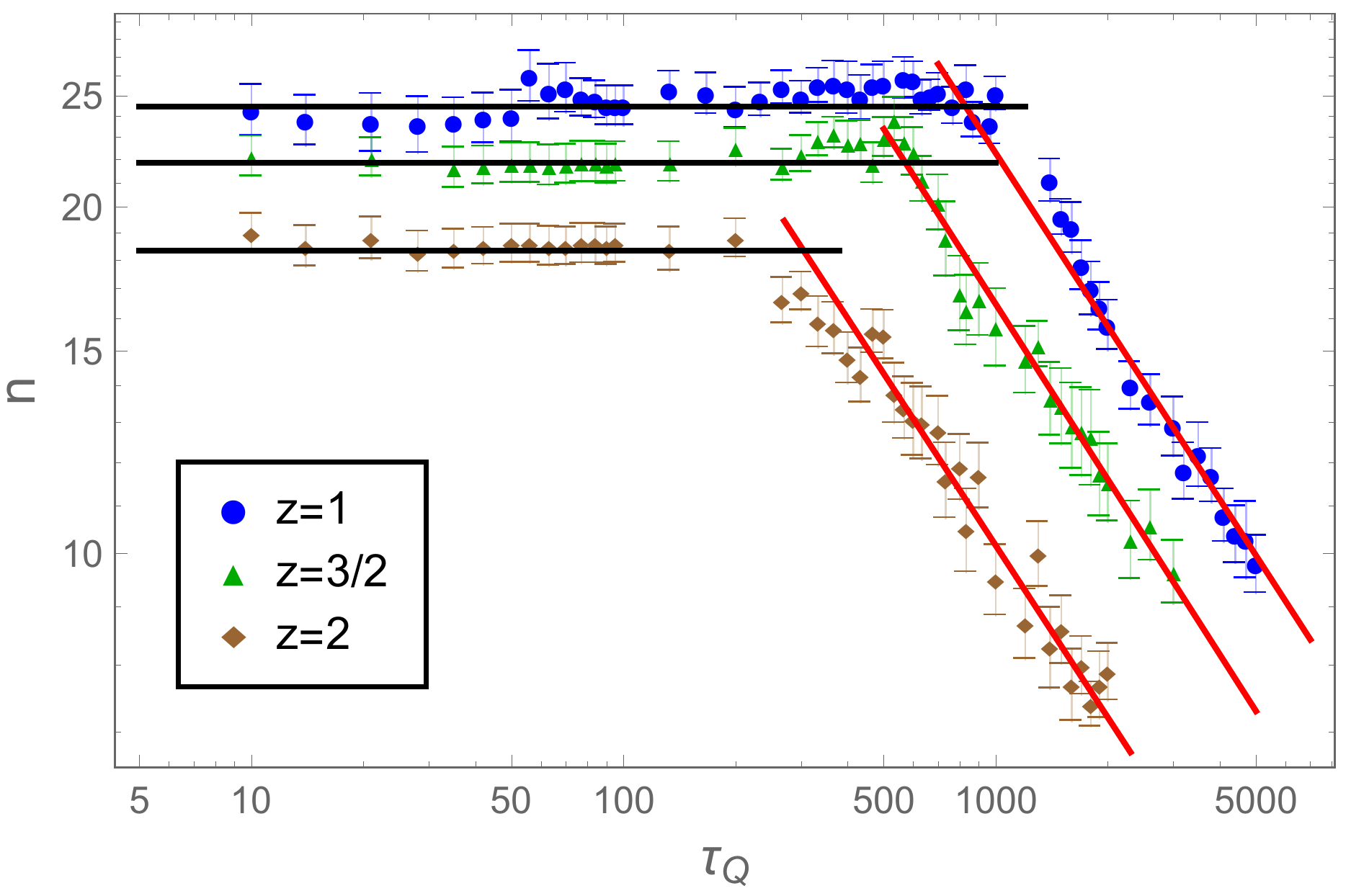}~~
\includegraphics[trim=0cm 0cm 0cm 0cm, clip=true, scale=0.38,  angle=0] {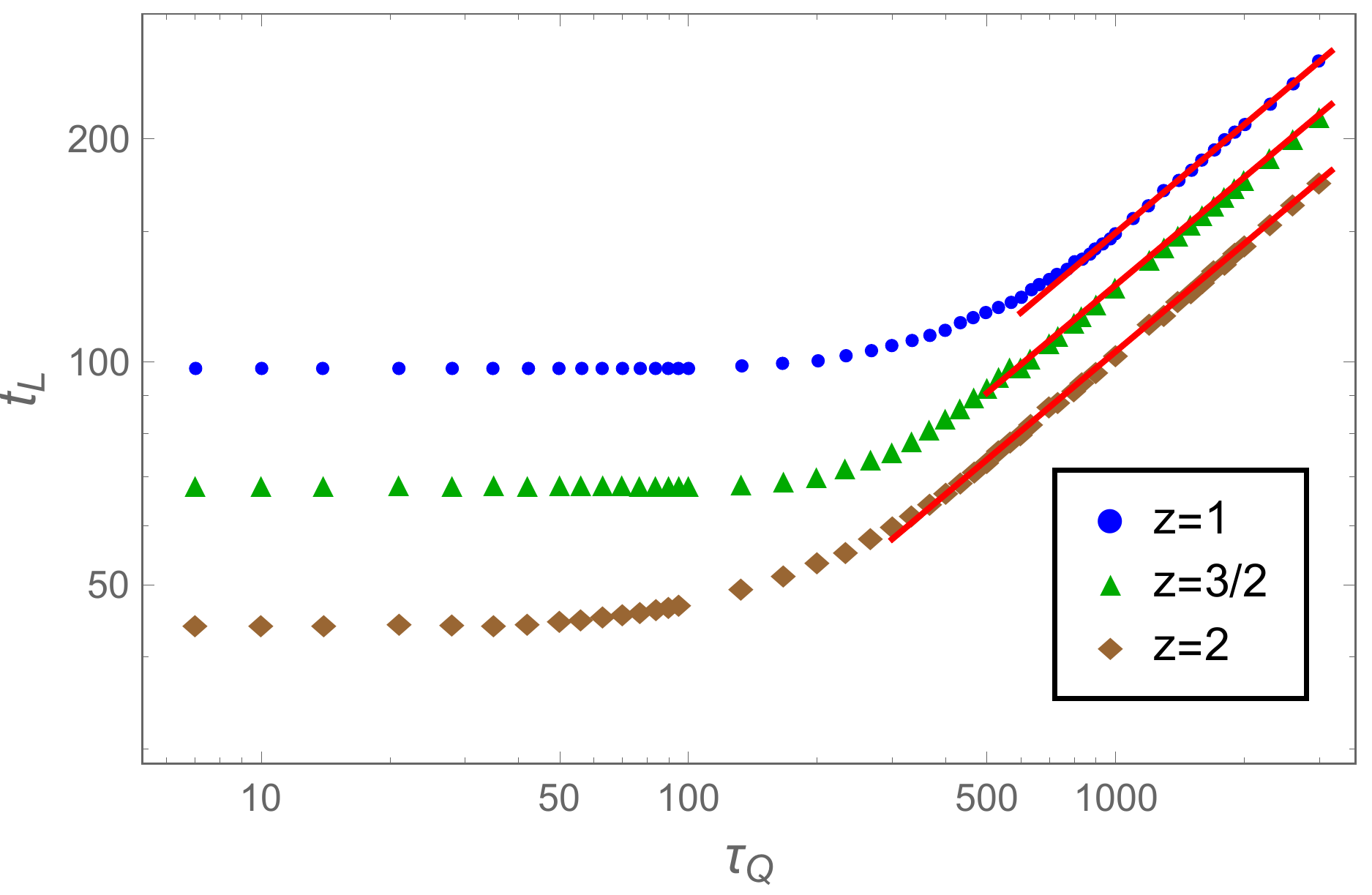}
\caption{\label{NT}
 (Left panel) Double logarithmic relation between vortex number density $n$ and quench time $\tau_Q$ for Lifshitz exponents {{$z=(1, 3/2,2)$. The dots, triangles and diamonds }}are the numerical data while the straight lines are the fitting lines.  For the slow quench (bigger $\tau_Q$ and red lines), both scaling relations satisfy the KZ scaling laws very well. However, for fast quench (smaller $\tau_Q$ and blue lines), vortex number density $n$ is almost constant beyond the scope of KZM. (Right panel) Double logarithmic relation between the lag time $t_L$ and quench time $\tau_Q$. For slow quench the relation between $t_L$ and $\tau_Q$ satisfy the KZ scalings very well, while for fast quench $t_L$ keeps almost constant. For both panels, the final equilibrium temperature is $T=0.8T_c$ and the size of the boundary system is $(x,y)=(50,50)$.}
\end{figure}

\subsection{Vortex number density and ``freeze-out'' time}
\label{sec:density}
In the left panel of Fig.\ref{NT}, we show the relation between the number density of vortices $n$ and quench time $\tau_Q$ under different Lifshitz exponents $z$. $n$ was counted in the final equilibrium state. For $z=1$, vortex numbers are almost the same ($n\approx25$) in the fast quench regime, which is consistent with previous results in condensed matter or holography \cite{Chesler:2014gya,Sonner:2014tca,Zeng:2019yhi}. {{For $z=(3/2,2)$, vortex number density $n\approx(22,18)$ in the fast quench regime. However, for slower quench, the scaling laws between $n$ and $\tau_Q$ for $z=1$ is roughly $n =n_1 \tau_Q^a$ with $n_1\approx(703.9170\pm1.1333)$ and $a\approx (-0.4998\pm0.0162)$, in which the error bars stand for the standard deviations. For $z=3/2$ and $z=2$, the relations are $n\approx (482.62\pm1.1431)\times\tau_{Q}^{(-0.4897\pm0.0183)}$ and $n\approx (312.9211\pm1.2379)\times\tau_{Q}^{(-0.4961\pm0.0315)}$ respectively.}}

As we have stated in the previous subsection, the lag time $t_L$ that defined as order parameter begins to grow rapidly can reflect the ``freeze-out'' time $\hat t$ \cite{Das:2011cx,Sonner:2014tca}. In numerics we operationally set $t_L$ as $\langle O\rangle\sim0.1$ following \cite{Sonner:2014tca,Zeng:2019yhi,Das:2011cx}.
On the right panel of Fig.\ref{NT} we exhibit the relation between $t_L$ and $\tau_Q$. The error bars are not shown, since they are very tiny. We see that for fast quench the lag time is almost constant for $z=(1, 3/2,2)$. However, for slow quench, one can read that $t_{L}\approx5.0540\times\tau_{Q}^{0.4893}$ for the Lifshitz exponent $z=1$, {{$t_{L}\approx4.5001\times\tau_{Q}^{0.4834}$ for $z=3/2$ and $t_{L}\approx3.6064\times\tau_{Q}^{0.4845}$ for $z=2$.  Therefore, from the two scaling relations in Eqs.\eqref{eq:tfreeze} and \eqref{density}, one can readily evaluate the dynamic critical exponent $z_{\rm d}$ and the static critical exponent $\nu$ on the boundary as ($z_{\rm d}\approx 1.9579$, $\nu\approx0.4893$) for $z=1$, ($z_{\rm d}\approx 1.9743$, $\nu\approx 0.4740$) for $z=3/2$ and ($z_{\rm d}\approx1.9532$, $\nu\approx0.4812$) for $z=2$.}} The holographic results for $z_{\rm d}$ and $\nu$ are very close to the mean-field theory values with $z_{\rm d}=2$ and $\nu=1/2$.\footnote{From quasi-normal modes analysis, one can get the same results of mean-field theory with $z_{\rm d}=2$ and $\nu=1/2$ \cite{Sonner:2014tca,Zeng:2019yhi}. It is not surprising, since the boundary field theory in the AdS/CFT is a mean-field theory in large $N_c$ limit \cite{Zaanen:2015oix}. }
Therefore, we see that the dynamic critical exponent $z_{\rm d}$ (as well as $\nu$) on the boundary is irrespective of the bulk Lifshitz exponent $z$.

\section{Conclusions and discussions}
\label{sec:conclusion}
We investigated the spontaneous formation and time evolution of topological defects from KZM in Lifshitz holography. The magnetic fluxes were found to be quantized and belonged to the type II superconductor. From the time evolution of the average condensate, we extracted the values of the lag time, which could reflect the ``freeze-out'' time. The KZ scaling relations, i.e., vortex number density to quench time and the lag time to quench time matched KZM very well. These two scaling relations implied that the dynamic critical exponent on the boundary field theory was irrespective of the Lifshitz exponent in the bulk. The reason is that from the AdS/CFT correspondence \cite{Zaanen:2015oix}, the boundary field theory is a mean-field theory in the large $N_c$ limit. This means whether the bulk is relativistic $(z=1)$ or non-relativistic $(z\neq1)$, the dynamic and static critical exponents in the boundary remain the same as those of mean-field theory. The topological defects (vortices in our paper) are formed due to the Kibble-Zurek mechanism in the boundary, thus, the powers of the scaling laws Eqs.\eqref{eq:tfreeze} and \eqref{density} will not change since there $\nu$ and $z_d$ are all the exponents in the boundary. This conclusion was also in line with previous discussions in \cite{Natsuume:2018yrg,Evans:2010np}, in which the authors perturbed the fields around the critical point to study the critical exponents. In our paper, without any perturbations, we saw that by directly studying the formations of topological defects, we arrived at the similar results.  According to discussions in \cite{hohenberg,Natsuume:2018yrg}, at least at finite temperature, critical dynamics is governed by the dynamics of the critical point itself rather than by the Lifshitz exponent $z$ in the underlying geometry. Therefore, it will be interesting to study the critical dynamics at zero temperature in Lifshitz geometry. We leave it for future work.

\acknowledgments
The authors are grateful for Wojciech H. Zurek's helpful discussions. This work was supported by the National Natural Science Foundation of China (Grants No. 11675140, 11705005 and 11875095).

\end{document}